\documentclass[11pt]{article} 
\usepackage{amssymb}
\usepackage{amsmath}
\usepackage{graphicx}
\usepackage{bm}
\newcommand{\mathsym}[1]{{}}

\usepackage{graphicx}
\usepackage{rotating}
\usepackage{color}
\usepackage{cancel}
\usepackage{hyperref}
\usepackage{etoolbox}
\usepackage{orcidlink}

\usepackage{doi}
\newcommand{\bra}{\begin{array}}
\newcommand{\era}{\end{array}}
\newcommand{\beq}{\begin{equation}}
\newcommand{\eeq}{\end{equation}}
\newcommand{\beqar}{\begin{eqnarray}}
\newcommand{\eeqar}{\end{eqnarray}}

\newcommand{\be}{\begin{equation}}
\newcommand{\ee}{\end{equation}}
\newcommand{\bea}{\begin{eqnarray}}
\newcommand{\eea}{\end{eqnarray}}
\newcommand{\bd}{\begin{displaymath}}
\newcommand{\ed}{\end{displaymath}}
\newcommand{\h}{\hbar}

\newcommand{\lb }{ \left (}
\newcommand{\rb }{ \right )}

\newcommand{\p }{ \partial}
\newcommand{\hh }{ \hat{H}}

\newcommand{\ep}{\epsilon}
\newcommand{\al }{\alpha}

\newcommand{\xh }{ \hat{x}}

\hypersetup{colorlinks=true,linkcolor=blue,urlcolor=blue,citecolor=blue}

\usepackage{cite}

\begin{document}

\vspace{20pt}

\begin{center}

{\LARGE \bf A new time-dependent quantum theory based on Tsallis' distribution

%\author{L.M. Nieto}% %
%\email{luismiguel.nieto.calzada@uva.es}
%\affiliation{Departamento de F\'{\i}sica Te\'orica, At\'omica y Optica and Laboratory for Disruptive \\ Interdisciplinary Science (LaDIS), Universidad de Valladolid, 47011 Valladolid, Spain}

\medskip
 }
\vspace{15pt}

{  Won Sang Chung\orcidlink{0000-0002-1358-6384}${}^{\dag}$\footnote{\href{mailto:mimip44@naver.com}{mimip44@naver.com}}, Georg Junker\orcidlink{0000-0003-2054-0453}${}^{\ddag\S}$\footnote{\href{mailto:georg.junker@fau.de}{georg.junker@fau.de},}, Luis M. Nieto\orcidlink{0000-0002-2849-2647}${}^{\ast}$\footnote{\href{mailto:luismiguel.nieto.calzada@uva.es}{luismiguel.nieto.calzada@uva.es}}, and Hassan Hassanabadi\orcidlink{0000-0001-7487-6898}${}^{\ast}$\footnote{\href{mailto:hha1349@gmail.com}{hha1349@gmail.com}}
}

\vspace{15pt}
{\sl ${}^{\dag}$Department of Physics and Research Institute of Natural Science,\\
 College of Natural Science, Gyeongsang National University, Jinju 660-701, Korea}\\

{\sl ${}^{\ddag}$Institut f\"{u}r Theoretische Physik I, Friedrich-Alexander Universit\"{a}t Erlangen--N\"{u}rnberg,
Staudtstra\ss e 7, 91058 Erlangen, Germany
 }

{\sl ${}^{\S}$European Southern Observatory, Karl-Schwarzschild-Stra\ss e 2, 85748 Garching, Germany
 }

{\sl ${}^{\ast}$Departamento de F\'{\i}sica Te\'orica, At\'omica y Optica and Laboratory for Disruptive \\ Interdisciplinary Science (LaDIS), Universidad de Valladolid, 47011 Valladolid, Spain
 }

%\vspace{8pt}
%E-mails:  {$
%{}^{\dag}$\href{mailto:mimip44@naver.com}{mimip44@naver.com},
%${}^{\ddag}$\href{mailto:georg.junker@fau.de}{georg.junker@fau.de},
%${}^{\ast}$\href{mailto:luismiguel.nieto.calzada@uva.es}{luismiguel.nieto.calzada@uva.es},
%${}^{\S}$\href{mailto:hha1349@gmail.com}{hha1349@gmail.com}}
\end{center}

\begin{abstract}
In this paper, inspired by Tsallis' probability distribution based on a $q$-deformed Boltzmann factor, we stipulate a new $q$-deformed quantum dynamics by applying the inverse Wick rotation  $ \beta \rightarrow i t$ to the Tsallis-deformed Boltzmann factor. We obtain a new time-dependent $q$-deformed Schr\"odinger equation. The free time-evolution of a Gaussian wave packet and that induced by an harmonic interaction are studied within this $q$-deformed quantum mechanical framework.
\end{abstract}

%\today

 \section{Introduction}
The year 2025 has been declared as the ``International Year of Quantum Science and Technology" by the United Nations General Assembly \cite{UNESCO}. It is celebrating 100 years of modern quantum mechanics, which started in 1925 with Heisenberg's fundamental paper \cite{Heis1925}. Since then quantum mechanics has not only formed our physical understanding of the micro cosmos, it has entered our daily live in numerous ways. Quantum technologies are nowadays almost everywhere in place, may it be lasers, electronics, mobile phones or medical diagnostic to just mention a few. One of the pillars of quantum mechanics has been Heisenberg's uncertainly relation discussed for the first time in his 1927 paper \cite{Heis1927} and being one of the basic foundations of the Copenhagen interpretation of quantum mechanics.

It is this Heisenberg algebra $[\hat{p},\hat{q}]=i\hbar$, i.e.\ the commutator between momentum operator $\hat{p}$ and position operator $\hat{q}$, and in particular its various deformations, which had attracted much attention among the physics community. For example, in 1947 Snyder \cite{Snyder} investigated the possibility of a Lorentz invariant discrete space-time and found that the Heisenberg's algebra requires an addition term on the right-hand-side (RHS) being proportional to $\hat{p}^2$. This deformation is still actively discussed in the literature in connection with quantum gravity \cite{Maggiore} and so-called gravitational quantum mechanics \cite{Nozari}. In 1950 Wigner \cite{Wigner} triggered another deformation of the Heisenberg algebra, where on the RHS now the parity operator appears instead. This was independently introduced by Dunkl \cite{Dunkl} and the associated deformed momentum operators are now well-established under his name, namely as  Dunkl operators. Another deformation of Heisenberg's algebra is based on so-called quantum groups an example of which is the $q$-deformed harmonic oscillator algebra $[a,a^\dag]_q=aa^\dag-qa^\dag a =1$ introduced in \cite{Arik,Macfarlane,Biedenharn}.

In quantum mechanics, the time-evolution of an initial quantum state function is obtained by acting the evolution operator on it. The evolution operator is unitary and, for a Hermitian time-independent Hamiltonian $\hh$, is given by $e^{ - it \hh}$. From now on and  throughout this paper we set $\h=1$.
In the thermal field theory \cite{Bellac,Kapusta,Das}, the Boltzmann factor $e^{-\beta \hh}$  is converted into the quantum evolution operator by transforming $ \beta \rightarrow i t$, that is, by applying the inverse Wick rotation. Here $\beta=1/(k_B T)$ stands for the inverse temperature  multiplied by Boltzmann's constant $k_B$. The probability of finding a quantum system in a state with energy $E$ is represented by that factor, $W_{E}=e^{-\beta E}/Z$, where the partition function basically serves as a normalisation factor $Z= {\rm Tr\,}e^{-\beta \hh}$. That is, the Boltzmann factor is the basis of statistics physics. In 1988 Tallis \cite{Tsallis} proposed a generalization of Boltzmann-Gibbs statistics by introducing a $q$-deformed Boltzmann factor of the form
\be\label{Boltzmann}
e_q ( - \beta E ) := [ 1- (1-q)\beta E ]^{\frac{1}{1-q}}\,,
\ee
also known as Tsallis’s distribution. It is this deformation which has revolutionised statistical physics since then \cite{Tsallis2009}. For example, the non-additivity of the corresponding Tsallis entropy has opened  new pathways to study systems being out of equilibrium or strongly correlated systems. For a nice introduction and overview with applications see, for example, the article by Cartwright \cite{Cartwright}. Here we only remark that in  \eqref{Boltzmann} the ordinary  Boltzmann factor is recovered in the limit $q \rightarrow 1$.

A first quantum mechanical investigation based on $q$-deformed plane waves is due to Nobre, Rego-Monteiro and Tsallis \cite{Nobre2011} who investigated associated Schrödinger, Klein-Gordon and Dirac equations. Here we adopt a different approach by considering the $q$-deformed Boltzmann factor \eqref{Boltzmann} and using it to postulate a $q$-deformed evolution operator via the inverse Wick transformation $ \beta \rightarrow i t$. This results in a new time-dependent quantum mechanical framework. Based on such $q$-deformed quantum evolution, we investigate the time-evolution of a free Gaussian wave packet and that of a Gaussian oscillating wave packet. 

The structure of the paper is as follows. In Section ~\ref{sec2}, starting from the $q$-deformed Tsallis-Boltzmann factor, we introduce a $q$-evolution operator and the associated time-dependent $q$-deformed Schrödinger equation. To show how the new formalism is applied, in Section ~\ref{sec3} we analyze the free $q$-evolution of a Gaussian wave packet, and in Section ~\ref{sec4} we study the $q$-evolution of the same Gaussian wave packet, but subjected to a harmonic oscillator-type potential. The paper ends with conclusions and some final remarks.

\section{$q$-evolution operator and time-dependent $q$-deformed Schr\"odinger equation}\label{sec2}

Let us start by considering the standard Schr\"odinger Hamiltonian for a quantum particle with mass $m>0$, which is of the form
\be\label{H}
\hh :=-\frac{1}{2m}\partial^2_x + V(x),
\ee
being $V(x)$ the external potential interacting with the particle. For simplicity we work in one dimension and the quantum states $\psi$ on which $\hh$ acts are living in the Hilbert space ${\cal H}=L^2(\mathbb{R})$.

The $q$-deformed Tsallis Boltzmann factor given in the equation \eqref{Boltzmann} suggests us to propose a new $q$-deformed evolution operator defined by $e_q ( - i t \hh )$. However, since in general this operator thus defined would not be unitary, a further modification is required, the correct definition of the $q$-deformed evolution operator being the following:
\be\label{Udef}
\hat{U}_q (t) := \frac{ e_q ( - i t \hh )}{|e_q ( - i t \hh )|}\,,
\ee
where the meaning of the operator in the denominator is
\be\label{eq4}
|e_q ( - i t \hh )| =\left( e_q ( - i t \hh )\ e_q ( + i t \hh ) \right)^{1/2}
\ee

From \eqref{Boltzmann} it is easy to obtain
\be
e_q (-iz) = [ 1 + (1-q)^2 z^2 ]^{\frac{1}{2(1-q)}} \exp \left\{ - i\,\frac{ \arctan [ (1-q) z]}{1-q} \right\}, \ee
where $\arctan$ stands for the principle branch of the inverse of the tangent function, and from \eqref{eq4}
we find
\be
|e_q ( - i t \hh )|  =  [ 1 + (1-q)^2 (t\hh)^2 ]^{\frac{1}{2(1-q)}}\,.
\ee
Thus, the unitary $q$-deformed evolution operator induced by Tsallis' $q$-deformed Boltzmann factor, reads
\be\label{U2}
\hat{U}_q (t) = \exp \left\{ - i\,\frac{ \arctan [ \varepsilon t\hh]}{\varepsilon} \right\}\,.
\ee
Here we have introduced the alternative deformation parameter $\varepsilon = 1-q$ with $\varepsilon \to 0$ being the undeformed quantum.
Any initial state $\psi(x, 0)\in{\cal H}$ then evolves in time according to
\be
\psi(x, t) = \hat{U}_q (t) \psi(x, 0)\,.
\ee
At this stage let us look at the semi-group property of $\hat{U}_q (t)$, respectively the Chapman-Kolmogorov relation, by considering
\be
\hat{U}_q (t_1+t_2)= \frac{ e_q ( - i (t_1+t_2) \hh )}{|e_q ( - i (t_1+t_2) \hh )|} = \frac{ e_q ( - i t_1 \hh )\otimes_q e_q ( - i t_2 \hh )}{1+\varepsilon^2(t_1+t_2)^2 \hh^2 }\,,
\ee
where we have utilised the $q$-product as defined in \cite{Nivanen2003,Borges2004,Tsallis2009}
\be\label{qprod}
a\otimes_q b := \left(a^\varepsilon + b^\varepsilon - 1\right)^\frac{1}{\varepsilon}
\ee
obeying the relation $e_q(x)\otimes_q e_q(y)=e_q(x+y)$. That is, for the composition law of the time evolution operator we need to first apply the $q$-product to the non-normalised operators $e_q(-i t\hh)$ and normalise the result afterwards to achieve unitarity.

The corresponding time-dependent $q$-deformed Schr\"odinger equation then reads
\be
i \frac{\p}{\p t} \psi (x, t) = \hh [ 1+ \varepsilon^2 t^2 \hh^2 ]^{-1} \psi(x, t)\,.
\ee
In other words, we have found a new effective time-dependent Hamiltonian induced by the original one (2) as follows
\be
 \hh_{\rm eff}(t):= \frac{\hh}{1+ \varepsilon^2 t^2 \hh^2 }\,.
\ee
For small $\varepsilon t$ the effective Hamiltonian reads
\be
\hh_{\rm eff}(t)= \hh  - \left(\varepsilon t\right)^{2} \hh^3 +O\left((\varepsilon t)^4\right)
\ee
and indicates that the limit of vanishing deformation parameter $\varepsilon$ is basically that same as for short times $t$.
Hence, we expect that the time evolution of the deformed quantum system deviates from the standard behaviour with a leading term of the order
$(\varepsilon t)^2$.

We may also look at the behaviour for $t\to\infty$. Assuming that our original Hamiltonian \eqref{H} is bounded from below by $\hh >0$ we observe that
\be
\lim_{t\to\infty}\hat{U}_q (t)= e^{-\frac{i\pi}{2\varepsilon}}\,.
\ee
That is, for large $t$ the dynamics in essence stops and the final state (i.e.\ $t\to \infty$) is up to a trivial phase factor identical to the original state at time $t=0$,
\be
\lim_{t\to\infty} \psi(x, t) = e^{-\frac{i\pi}{2\varepsilon}}\psi(x, 0)\,.
\ee
Note, when looking at the behaviour for small $\varepsilon$ we must take into account that this may only be valid when $t$ is not too large.

As in standard quantum mechanics, the probability of finding a particle at position $x$ at time $t$ is given by
\be
P(x, t) =|\psi(x, t)|^2\,.
\ee
Similarly, the expectation values of an operator $\hat{A}$ in a quantum state described by $\psi(x,t)$ reads
\be
\langle \hat{A} \rangle =  \int_{-\infty}^{\infty} \psi^*(x, t) {\hat A} \psi(x, t) dx \,.
\ee
We will make use of that when discussing the spreading of a Gaussian wave packet in the absence of an external potential in the next section.

\section{Time evolution of Gaussian wave packet}\label{sec3}

In this section we will investigate the time evolution of a free Gaussian wave packet. Hence, we set $V=0$ in \eqref{H} and assume as initial state the normalized Gaussian wave function
\be\label{Gausspsi0}
\psi(x, 0) = \left[\frac{ 1}{ \sqrt{2 \pi\sigma^2}} \right]^{1/2} e^{ - \frac{x^2}{4 \sigma^2}}.
\ee
Obviously the probability density of finding a particle at position $x$ at time $0$ is then given by the centralized Gaussian distribution
\be\label{GaussP0}
P(x, 0) = |\psi(x, 0)|^2 = \frac{ 1}{ \sqrt{2 \pi\sigma^2}}\, e^{ - \frac{x^2}{2 \sigma^2}}
\ee
with a vanishing expectation value of position, $\langle x\rangle =0$, and a width characterised by $\sigma$ with $\langle x^2 \rangle = \sigma^2$.

Passage from position to momentum representation is accomplished via a Fourier transformation resulting in another Gaussian in momentum space
\be\label{Gaussphi0}
\phi(k, 0) = \frac{1}{\sqrt{2\pi}} \int_{-\infty}^{\infty} e^{ - i k x } \, \psi(x, 0)\,  dx
= \left[ \frac{2 \sigma}{\sqrt{2 \pi}}\right]^{1/2} e^{ - \frac{1}{4} ( 2 \sigma k )^2}.
\ee
Let us note that a plane wave $e^{i kx}$ is an eigenfunction of the momentum operator  $\hat{P}=-i\partial_x $ with wave number $k\in \mathbb{R}$ being the corresponding eigenvalue, i.e.\ $\hat{P}e^{i kx}=k e^{i kx}$. In turn, they are also eigenfunctions of the free undeformed Hamiltonian $\hh = \frac{\hat{P}^2}{2m}$ with eigenvalues given by $E_k = \frac{ k^2}{2m}$.
Therefore, the time evolution of such a plane wave $e^{i kx}$ generated by the deformed time evolution operator \eqref{U2} results in a multiplicative phase factor for each $k$. Hence, the time-evolved state, $\psi(x, t)$, is given by the inverse Fourier transform of the time-evolved momentum distribution
\be
\phi(k, t)= \hat{U}_q (t)\phi(k,0)= \exp \left\{ - \frac{i}{\varepsilon}\arctan \left[ \frac{\varepsilon t k^2}{2m}\right] \right\}\phi(k,0)\,.
\ee
Explicitly we have
\be
\psi(x, t) = \frac{1}{\sqrt{2\pi}} \left[ \frac{2 \sigma}{\sqrt{2 \pi}}\right]^{1/2}
\int_{-\infty}^{\infty} e^{- \sigma^2 k^2} e^{i kx}   \exp\left\{ - \frac{i}{\varepsilon} \arctan \left[ \frac{\varepsilon tk^2}{2m}\right]\right\}dk
\ee
This integral does not give a closed expression. So we consider the case of small $\varepsilon$. Up to second order in $\varepsilon$, we have
\be
\psi(x, t) = \frac{1}{\sqrt{2\pi}} \left[ \frac{2 \sigma}{\sqrt{2 \pi}}\right]^{1/2}
\int_{-\infty}^{\infty} e^{- \sigma^2 k^2} e^{i kx}  e^{- \frac{ i k^2 t}{ 2m}} \left[ 1 + i \varepsilon^2 \frac{k^6 t^3}{24m^3} \right] dk +O(\varepsilon^4)
\ee
which, when explicitly integrated, results in
\be
\psi(x, t) = \sqrt{ \frac{1} { \sigma \sqrt{2 \pi}\lb  1 + \frac{it}{\tau}\rb}}
e^{ - \frac{x^2}{ 4 \sigma^2 \lb 1 + \frac{it}{\tau}\rb}}
\lb 1 + i \frac{\varepsilon^2t^2}{\tau^2} F(t, x ) +O(\varepsilon^4)\rb
\ee
where we have introduced the time scale  $\tau = 2m \sigma^2$. The remaining factor $F$ is given by
\be\label{F}
F(t, x)
= \frac{1}{24} \frac{1}{ \lb 1 + \frac{it}{\tau}\rb^3}
\left[ 15 - \frac {45}{2} \frac{1}{  1 + \frac{it}{\tau}} \lb \frac{x}{\sigma}\rb^2
+ \frac {15}{4} \frac{1}{  \lb 1 + \frac{it}{\tau}\rb^2 } \lb \frac{x}{\sigma}\rb^4
- \frac {1}{8} \frac{1}{  \lb 1 + \frac{it}{\tau}\rb^3 } \lb \frac{x}{\sigma}\rb^6 \right].
\ee
Here we observe the previously expected result that the leading order for small $\varepsilon$, respectively small $t$, should be given by a term proportional to $(\varepsilon t)^2$.
The probability density  is then given by
\be
P(x, t) = \frac{1} { \sigma \sqrt{2 \pi}\sqrt{  1 + \frac{t^2}{\tau^2}}}e^{ - \frac{x^2}{ 2 \sigma^2 \lb  1 + \frac{t^2}{\tau^2}\rb}} \left[ 1 - \varepsilon^2 \lb \frac{t}{\tau}\rb^3 g(x, t) + O(\varepsilon^4)\right]
\ee
where
\be
g(x, t) = \textstyle A(\frac{t}{\tau}) + B(\frac{t}{\tau})\,  \lb \frac{x}{\sigma}\rb^2 + C(\frac{t}{\tau})\, \lb \frac{x}{\sigma}\rb^4 + D(\frac{t}{\tau})\, \lb \frac{x}{\sigma}\rb^6
\ee
with time dependent coefficients
\be
\begin{array}{ll}
\displaystyle
A(z)= \frac{5}{4} \frac {z^2 -3}{ \lb  1 + z^2\rb^3}\, ,\qquad &\displaystyle B(z)= - \frac{15}{2} \frac { z^2 -1}{ \lb  1 + z^2\rb^4}\,,\\[4mm]
\displaystyle
C(z)=  \frac{5}{16}  \frac{ 5- 10z^2 + z^4}{ \lb  1 + z^2\rb^5}\,,\qquad &\displaystyle D(z)=  \frac{1}{48}  \frac{3-10z^2+3z^4} {\lb  1 + z^2\rb^6}\,.
\end{array}
\ee

\begin{figure}[htb]
      	\centering{
\includegraphics[width=7.5cm]{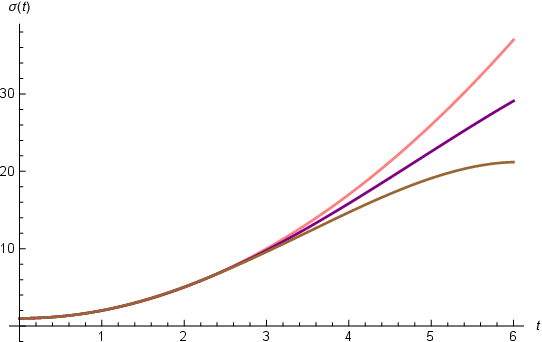}
}
\caption{\footnotesize Plot of the spreading \eqref{sigmat}  of a freely evolving Gaussian wave packet with initial width $\sigma =1$ and time scale $\tau=1$, for $ \varepsilon^2=0$ (pink), $\varepsilon^2=0.0005$ (purple) and $ \varepsilon^2=0.001$ (brown).}
\label{fig1}
\end{figure}

Now we are in a position to evaluate the spreading of the wave packet up to second order in $\varepsilon$, that is,
\be\label{sigmat}
\langle \xh^2 \rangle = \sigma(t)^2
=\sigma^2 \lb  1 + \frac{t^2}{\tau^2}\rb
\left[ 1- 15 \varepsilon^2 \lb \frac { \frac{t}{\tau}}{ 1 + \frac{t^2}{\tau^2}} \rb^3 \lb
3- 6\frac{t^2}{\tau^2}+ \frac{t^4}{\tau^4} \rb  +O(\varepsilon^4)\right].
\ee
In Figure~\ref{fig1} we have plotted the spreading of the width $\sigma(t)$ for parameters  $\sigma =1 , \tau=1$ and various values of $\varepsilon$. Namely, for $ \varepsilon^2=0$ (pink), $ \varepsilon^2=0.0005$ (purple) and $ \varepsilon=0.001$ (brown). We clearly see the slowing down of spreading with increasing $\varepsilon$ as expected from above result.

Let us also briefly consider the time-dependent uncertainty relation by recalling $\Delta x=\sigma(t)$ and noting that $\Delta p = \sqrt{\langle k^2 \rangle}= 1/2\sigma$ remains constant, resulting in
\be
\Delta p \Delta x = \frac{1}{2}\,\sqrt{1 + \frac{t^2}{\tau^2}}\left[ 1- \frac{15}{2} \varepsilon^2 \lb \frac { \frac{t}{\tau}}{ 1 + \frac{t^2}{\tau^2}} \rb^3 \lb
3- 6\frac{t^2}{\tau^2}+ \frac{t^4}{\tau^4} \rb  +O(\varepsilon^4)\right].
\ee

\section{ Oscillating wave packet}\label{sec4}
As a second example we consider the modified quantum dynamics in the presence of an harmonic interaction. That is, we consider the undeformed system
\be\label{Homega}
\hh = -\frac{1}{2m}\partial^2_x + \frac{m}{2}\omega^2 x^2\,,\qquad \omega >0\,.
\ee
The eigenvalues and normalised eigenfunctions of \eqref{Homega} are well-known and given by
\be
E_n=\omega\left(n+\frac{1}{2}\right)\,,\qquad u_n (x) = \lb \frac{\al}{ \sqrt{\pi} 2^n n!} \rb^{1/2} H_n ( \xi) e^{ - \frac{1}{2} \xi^2 }\,,\qquad n=0,1,2,3,\ldots.
\ee
In the above $H_n$ stands for the Hermite polynomial of order $n$, $\xi = \al x$ and $\al = \sqrt{m\omega}$.

The deformed time evolution of such an eigenstate is then given via \eqref{U2} in the explicit form
\be
\hat{U}_q (t) u_n (x) = \exp \left\{ - i\,\frac{ \arctan [ \varepsilon \omega t\left(n+\frac{1}{2}\right)]}{\varepsilon} \right\}u_n (x)\,.
\ee
Therefore, decomposing an arbitrary initial state $\psi (x, 0)$ into the harmonic oscillator eigenstates,
\be
\psi (x, 0)=\sum_{n=0}^{\infty}A_n\,u_n (x) \,,\qquad A_n = \int_{-\infty}^{\infty} u_n (x)\psi (x, 0) dx\,,
\ee
we arrive at the quantum state at a given time $t>0$
\be\label{psixtHO}
\psi (x, t) = \sum_{n=0}^{\infty} A_n u_n (x) \exp \left\{ - i\,\frac{ \arctan [ \varepsilon \omega t\left(n+\frac{1}{2}\right)]}{\varepsilon}\right\}.
\ee
It is this sum which may be challenging to bring into a closed form expression.

Now let us consider the case where the initial state is given by a Gaussian centered around a fixed point $a$,
\be\label{psi0HO}
\psi(x, 0) = \left(\frac{\al^2}{\pi}\right)^{1/4} e^{ - \frac{\al^2 }{2} ( x - a)^2 }.
\ee
The corresponding amplitudes may be calculated explicitly and read
\be
A_n = \frac{\xi_0^n}{\sqrt{ 2^n n!}} e^{- \frac{1}{4} \xi_0^2}\,,\qquad\xi_0 = \al a\,.
\ee
At that stage we utilise the generating function for Hermite polynomials in the form
\be
F_k(\xi,p):= \lb p\frac{\partial} {\partial p}\rb^k  e^{ 2 p\xi -p^2 } = \sum_{n=0}^{ \infty} \frac{ H_n (\xi)}{n!} n^k p^n
\ee
and note
\be
\hskip-0.5cm
\begin{array}{l}
  F_0(\xi,p) = e^{ 2 p\xi - p^2 },\\
  F_1(\xi,p) = e^{ 2 p\xi - p^2 }(-2p^2 + 2p\xi),\\
  F_2(\xi,p) = e^{ 2 p\xi - p^2 }(4p^4-4p^2 - 8p^3\xi +4 p^2 \xi^2 + 2p\xi),\\
  F_3(\xi,p) = e^{ 2 p\xi - p^2 }(-8p^6 + 24 p^4 -8p^2 + 24p^5\xi -24p^4\xi^2 + 8p^3 \xi^3 - 36p^3 \xi + 8p^2\xi^2 + 2p\xi).
\end{array}
\ee
Considering only terms up to second order in $\varepsilon$ in \eqref{psixtHO} and setting $p=\frac{\xi_0}{2}e^{-i\omega t}$, we arrive at
\be
\psi(x, t)=  \left(\frac{\al^2}{\pi}\right)^{1/4} \exp \left[ -\frac{\xi^2}{2} -\frac{\xi_0^2}{4}
- \frac{ i \omega t}{2} + 2 p\xi  -p^2 \right] G(t, x)
 \ee
where
\be
G(t, x)= 1+ i  \frac{  \ep^2}{24}(wt)^3\left[1 +\frac{6\,F_1(\xi,p)}{e^{ 2 p\xi - p^2 }}+ \frac{12\, F_2(\xi,p)}{e^{ 2 p\xi - p^2 }} + \frac{8\, F_3(\xi,p)}{e^{ 2 p\xi - p^2 }}\right] +O(\varepsilon^4)\,.
\ee
Or more explicitly
\be
\begin{array}{l}
G(t, x)= 1+ i  \varepsilon ^2(wt)^3 \times \\
  \qquad \times \left[1- \frac{8}{3} p^6 +10p^4 - \frac{31}{6}p^2 +8p^5\xi-8p^4\xi^2+\frac{8}{3} p^3\xi^3 +16p^3\xi+ \frac{14}{3}p^2\xi^2+\frac{5}{3} p\xi  \right] +O(\varepsilon^4)\,.
\end{array}\ee
The associated probability density $P(x, t)= |\psi(x, t)|^2$ is of the form
\be\label{PHO}
P(x, t)= \sqrt{\frac{\al^2}{\pi}}e^{ -\al^2 ( x- a \cos \omega t)^2 }|G(x,t)|^2
\ee

\begin{figure}[htb]
      	\centering{
\includegraphics[width=7.4cm]{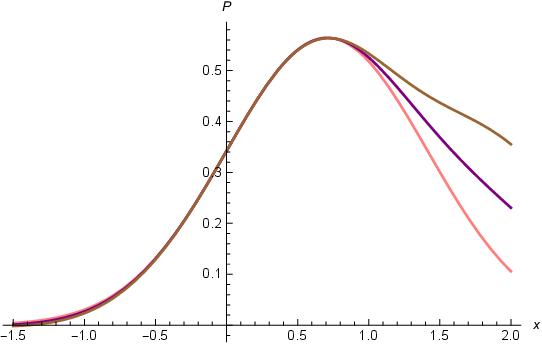}
\qquad\qquad
\includegraphics[width=7.4cm]{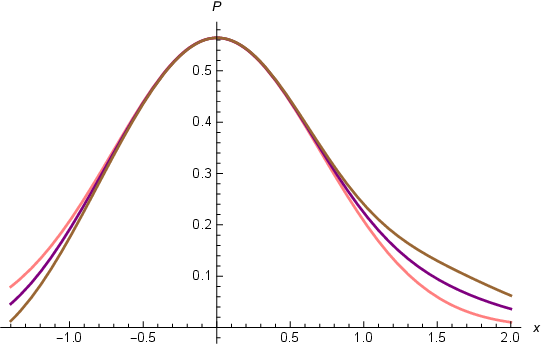}
}
\caption{\footnotesize Plot of the probability density \eqref{PHO}  for the oscillating wave packet at $t = \pi/4\omega $ (left panel) and  $\omega t = \pi/2\omega$ (right panel)  with parameters $a =1$ and $\al=1$, and for $\varepsilon^2=0$ (pink), $ \varepsilon^2=0.01$ (purple), and $ \varepsilon^2=0.02$ (brown).}
%\caption{Plot of $P(x, t)$ with $a =1 , \al=1$ for $ \varepsilon^2=0$ (pink), $ \varepsilon^2=0.01$ (purple) and $ \varepsilon^2=0.02$ (brown). Left panel $\omega t = \pi/4$, right panel $\omega t = \pi/2$.
\label{fig2}
\end{figure}

%{\red\small
%\noindent Little calculation to be removed
%$$
%|G(x,t)|^2 = |1+ i  \varepsilon^2(\omega t)^3\left[\cdots\right]|^2 = 1 -2\varepsilon^2 (\omega t)^3 {\rm %Im\,}[\cdots]+O(\varepsilon^4)\quad
%\mbox{and}\quad {\rm Im\,}p^n= -\frac{\xi_0^n}{2^n}\sin(n\omega t)
%$$
%$$
%\begin{array}{l}
%|G(t, x)|^2 = 1+ 2  \varepsilon ^2(\omega t)^3 \left[\frac{\xi_0^6}{24}\sin(6\omega t) -\frac{5\xi_0^4}{8} \sin(4\omega t)+\frac{31\xi_0^2}{24} \sin(2\omega t) - \frac{\xi_0^5\xi}{4} \sin(5\omega t)+ \right. \\
%\qquad\left.   + \frac{\xi_0^4\xi^2}{2}\sin(4\omega t) -\frac{\xi_0^3\xi^3}{3}\sin(3\omega t) -2\xi_0^3\xi\sin(3\omega t) -\frac{7\xi_0^2\xi^2}{6}\sin(2\omega t)
%                      -\frac{5\xi_0\xi}{6}\sin(\omega t) \right] +O(\varepsilon^4)\,.
%\end{array}
%$$
%Is there any need to calculate the maximum for $|P(x,t)|^2$ or is it sufficient to show Fig 2 and 3. Could both figures be redone with the new result?\\
%}

\noindent with
\be\label{G^2}
\begin{array}{rl}
|G(t, x)|^2 =& 1+ 2  \varepsilon ^2(\omega t)^3 \left[\frac{\xi_0^6}{24}\sin(6\omega t) - \frac{\xi_0^5\xi}{4} \sin(5\omega t) + \frac{\xi_0^4}{2}\left(\xi^2 -\frac{5}{4}\right)\sin(4\omega t)
 - \right. \\
\qquad&\left.  -\frac{\xi_0^3\xi}{3}\left( \xi^2+6\right)\sin(3\omega t) + \frac{\xi_0^2}{6}\left(\frac{31}{4}-7\xi^2\right) \sin(2\omega t)-\frac{5\xi_0\xi}{6}\sin(\omega t) \right] +O(\varepsilon^4)\,.
\end{array}
\ee
The probability density oscillates in time with a period $T =\frac{2\pi}{\omega}$. In the left panel of Figure~\ref{fig2} we show $P(x, T/8)$ for $a =1$,  $\al=1$ and various values for $\varepsilon$, 
namely $ \varepsilon^2=0$ (pink), $ \varepsilon^2=0.01$ (purple) and $ \varepsilon^2=0.02$ (brown); in the right panel of this figure the plot of $P(x, T/4)$ is shown.
We observe a behaviour being different to that of the freely evolving wave packet. Obviously, here the spreading of the wave packet behaves differently for the regions $x<0$ and $x>0$, which is due to the asymmetric initial wave packet \eqref{psi0HO}. 
 In the region $x>0$, in contrast to the free motion, the spreading increases with increasing deformation parameter $\varepsilon$. While for region $x<0$ we see the opposite behavior but much less dominant. We also observe an oscillatory behaviour in time for this spreading as expected due to the harmonic interaction. This is more pronounce for the region $x>0$.

%\begin{figure}
%\includegraphics[width=8cm]{a7-3.eps}
%\caption{Plot of $P(x, t)$ with $a =1 , \al=1, \omega t = \pi/2$ for $ \varepsilon^2=0$ (pink), $ \varepsilon^2=0.01$ (purple) and $ \varepsilon^2=0.02$ (brown).
% }
%\label{ex1}
%\end{figure}

\section{Conclusion}\label{sec5}

In this paper we introduced a new unitary $q$-deformed evolution operator generating a quantum mechanical time evolution, which may be called Tsallis evolution. This was done by transforming $ \beta \rightarrow i t$ in the $q$-deformed Boltzmann factor appearing in the non-extensive $q$-entropy theory of Tsallis. Using this we derived a time-dependent $q$-deformed Schr\"odinger equation. We investigated the time-evolution of a Gaussian wave packet based on this $q$-deformed Schr\"odinger-Tsallis equation. Because there is no closed expression for the time evolution of such wave packet, we considered the time-dependent Gaussian wave packet up to a second order in $\varepsilon = 1-q$. Thus, our results become physical for evolution during a small time as discussed in Section~\ref{sec2}. We observed that for the free Tsallis-deformed quantum dynamics the spreading of such a wave packet is increased by this deformation.
In a second example  we introduced an external harmonic interaction and assumed for the initial state a decentralized  Gaussian wave packet. Here the observed time-dependent behaviour of the spreading was rather different as that observed for the free deformed quantum dynamics, as discussed above. Clearly we require further investigation of the Tsallis evolution for additional quantum mechanical systems such as two-level Hamiltonian with the initial state being an overlap of the two eigenstates with different energies like, e.g., a Rabi model Hamiltonian. Another interesting possibility is to study the $q$-deformed evolution of a quantum state confined in a box. We are currently working on both lines of research.

Let us conclude that the Tsallis deformation of the quantum mechanical time evolution has an observable effect but may not be calculated in a closed form. Only for eigenfunctions of the Hamiltonian $\hh$, say $\hh \psi_E = E\psi_E$, we may find such a closed form:
\be
\psi_E(t)=\exp \left\{ - i\,\frac{ \arctan [ \varepsilon tE]}{\varepsilon}\right\}\psi_E\,.
\ee
But in this case, the effect will not materialize because $|\psi_E|^2=|\psi_E(t)|^2=constant$. Therefore, for any effect to be observed, any initial state must be a linear combination of several eigenstates, as described here for the free quantum system and the harmonic oscillator.

\section*{Acknowledgements}

The research of L.M.N. and H.H. was supported by the Q-CAYLE project, funded by the European Union-Next Generation UE/MICIU/Plan de Recuperacion, Transformacion y Resiliencia/Junta de Castilla y Leon (PRTRC17.11), and also by project PID2023-148409NB-I00, funded by MICIU/AEI/10.13039/501100011033. Financial support of the Department of Education of the Junta de Castilla y Leon and FEDER Funds is also gratefully acknowledged (Reference: CLU-2023-1-05).

\end{document}